\documentclass[runningheads]{llncs}

\usepackage{graphicx}
\usepackage{comment}
\usepackage{amsmath,amssymb}
\usepackage{color}
\usepackage{url}
\usepackage{hyperref}
\usepackage[capitalise, noabbrev]{cleveref}

\newif\ifreview
\reviewfalse

\ifreview
	\usepackage{lineno}

	\linenumbers
\fi

\begin{document}


\def\SubNumber{93}

\def\GCPRTrack{Track: Pattern recognition in the life and natural sciences}

\title{How Reliable Are Out-of-Distribution Generalization Methods for Medical Image Segmentation? 
    \thanks{
         Supported by the Bundesministerium für Gesundheit (BMG) with grant [ZMVI1-2520DAT03A]. The final authenticated version of this manuscript will be published in Lecture Notes in Pattern recognition in the life and natural sciences - DAGM GCPR 2021 (doi will follow).
    }
}
\titlerunning{How Reliable is OoD Generalization?}


\ifreview
\else

	\author{Antoine Sanner \orcidID{0000-0002-4917-9529} \and
	Camila González \orcidID{0000-0002-4510-7309} \and
	Anirban Mukhopadhyay \orcidID{0000-0003-0669-4018}}
	
	\authorrunning{Sanner et al.}

	\institute{Technical University of Darmstadt, Karolinenpl. 5, 64289 Darmstadt, Germany\\
	\email{\{firstname.lastname\}@GRIS.TU-DARMSTADT.DE}}
\fi

\maketitle              

\begin{abstract}

The recent achievements of Deep Learning rely on the test data being similar in distribution to the training data.
In an ideal case, Deep Learning models would
achieve \textbf{Out-of-Distribution} (\textbf{OoD}) \textbf{Generalization}, i.e. reliably make predictions on out-of-distribution data.
Yet in practice, models usually fail to generalize well when facing a shift in distribution. Several methods were thereby designed to improve the robustness of the features learned by a model through \textbf{Regularization-} or \textbf{Domain-Prediction-based} schemes.
Segmenting medical images such as MRIs of the hippocampus is essential for the diagnosis and treatment of neuropsychiatric disorders. But these brain images often suffer from distribution shift due to the patient's age and various pathologies affecting the shape of the organ.
In this work, we evaluate OoD Generalization solutions for the problem of hippocampus segmentation in MR data using both fully- and semi-supervised training. 
We find that no method performs reliably in all experiments. Only the \textbf{V-REx} loss stands out as it remains easy to tune, while it outperforms a standard U-Net in most cases. 

\end{abstract}

\section{Introduction}

Semantic segmentation of medical images is an important step in many clinical procedures. In particular, the segmentation of the hippocampus from MRI scans is essential for the diagnosis and treatment of neuropsychiatric disorders. Automated segmentation methods have improved vastly in the past years and now yield promising results in many medical imaging applications \cite{Litjens2017}. These methods can technically exploit the information contained in large datasets. However, no matter how large a training dataset is or how good the results on the in-distribution data are, methods may fail on \textbf{Out-of-Distribution} (\textbf{OoD}) data. \textbf{OoD Generalization} remains crucial for the reliability of deep neural networks, as insufficient generalization may vastly limit their implementation in practical applications. 

\textit{Distribution shifts} occur when the data at test time is different in distribution from the training data. In the context of hippocampus segmentation, the age of the patient \cite{XU20081003} and various pathologies \cite{carmo_hippocampus_2020} can affect the shape of the organ. Using a different scanner or simply using different acquisition parameters can also cause a distribution shift \cite{Castro2020}.

OoD Generalization alleviates this issue by training a model such that it generalizes well to new distributions at test time without requiring any further training. Several strategies exist to approach this. \textbf{Regularization-based} approaches enforce the learning of robust features across training datasets \cite{arjovsky_invariant_2020,krueger_out--distribution_2020}, which can reliably be used to produce an accurate prediction regardless of context. On the other hand, \textbf{Domain-Prediction-based} methods focus on harmonizing between domains \cite{Dinsdale2020.10.09.332973,ganin_unsupervised_2015} by including a domain predictor in the architecture. 

In this work, we perform a thorough evaluation of several state-of-the-art OoD Generalization methods for segmenting the hippocampus. We find that no method performs reliably across all experiments. Only the Regularization-based \textbf{V-REx} loss stands out as it remains easy to tune, while its worst results remains relatively good. 
\section{Methods}

The setting for Out-of-Distribution Generalization has been defined \cite{arjovsky_invariant_2020} as follows. Data is collected from multiple environments and the source of each data point is known. An \textit{environment} describes a set of conditions under which the data has been measured. One can for instance obtain a different environment by using a different scanner or studying a different group of patients. These environments contain spurious correlations due for instance to dataset biases. 

Only fully labeled datasets from a limited set of environments is available during training, and the goal is to learn a predictor, which performs well on all environments. As such, a model trained using this method will theoretically perform well on unseen but semantically related data.

\begin{figure}
    \includegraphics[width=\textwidth]{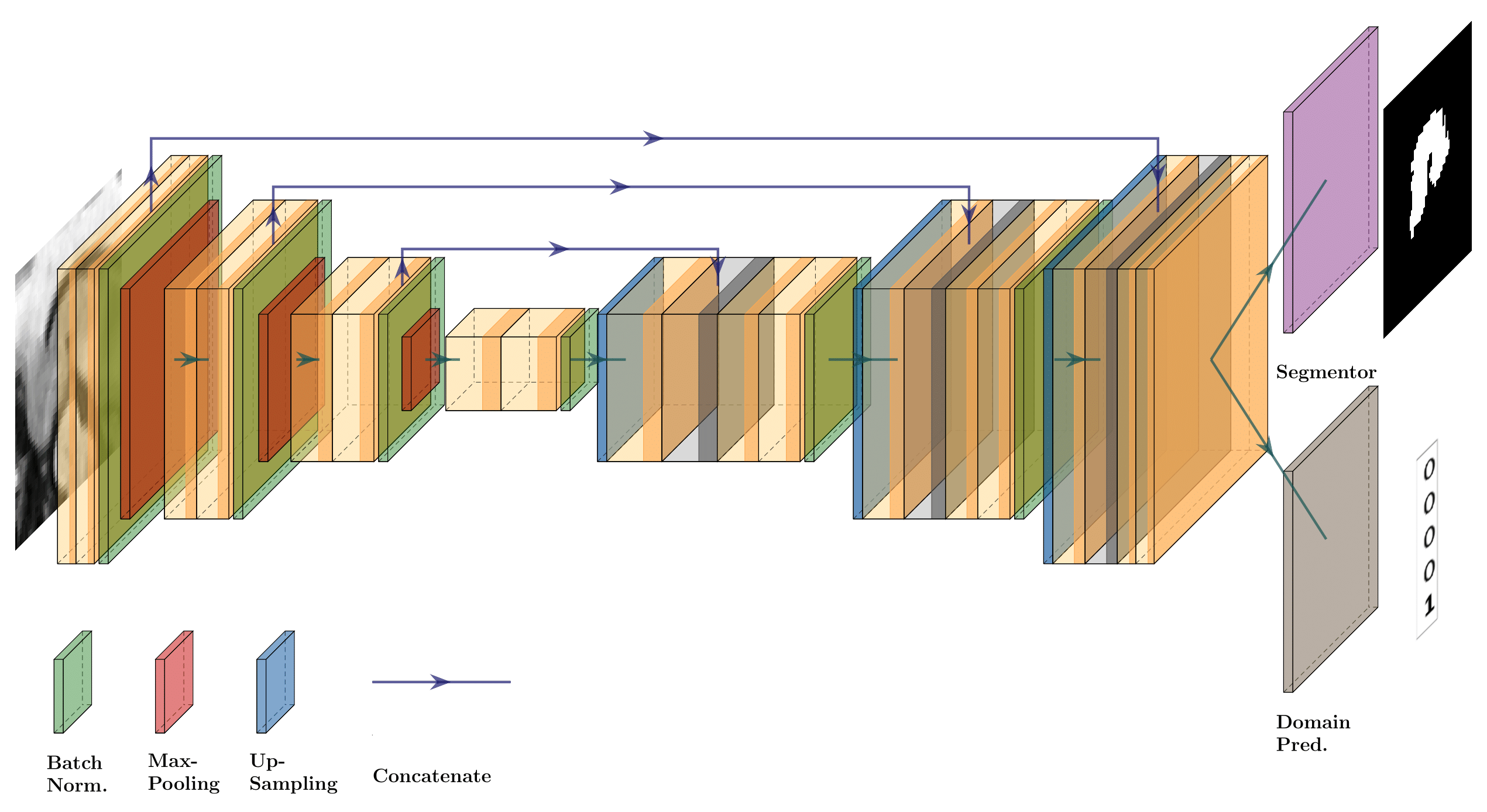}
    \caption{
Network architecture for the segmentation task with Domain-prediction mechanism. The U-Net generates features, which are then used by the segmentor for the main task and by the auxiliary head for domain prediction.
This schema contains some simplifications: the segmentor and domain predictor are implemented respectively using a convolutional layer and a CNN. The architecture displayed is 2D, while the implementation is 3D. 
    }
    \label{UNet_plot}
\end{figure}

\subsection{Regularization-based methods}
\label{oodg_methods}

Arjovsky et al. \cite{arjovsky_invariant_2020} formally define the problem of OoD Generalization as follows. Consider the datasets $D_e := \{(x_i^e, y_i^e)\}^{n_e}_{i=1}$ collected under multiple training environments $e \in \mathcal{E}_{tr}$. These environments describe the same pair of random variables measured under different conditions. The dataset $D_e$, from environment $e$, contains examples identically and independently distributed according to some probability distribution $P(X_e, Y_e)$. The goal is then to learn a predictor $Y \approx f(X)$, which performs well across all unseen related environments $\mathcal{E}_{all} \supset \mathcal{E}_{tr}$. The objective is to minimize:
\begin{equation*}
    R^{OoD}(f) = \max_{e \in \mathcal{E}_{all}} \mathcal{R}^e(f)
\end{equation*}
where $\mathcal{R}^e(f) := \mathbb{E}_{X^e, Y^e}[\ell(f(X^e), Y^e)]$ is the risk under environment $e$.

\textbf{IRMv1} \cite{arjovsky_invariant_2020} and Risk Extrapolation (\textbf{V-REx} and \textbf{MM-Rex}) \cite{krueger_out--distribution_2020} add regularization to the training loss to enforce strict equality between training risks by finding a data representation common to all environments. \textbf{IRM Games} \cite{ahuja_invariant_2020} poses the problem as finding the Nash equilibrium of an ensemble game.

More precisely, Krueger et al. \cite{krueger_out--distribution_2020}  propose to solve this problem by minimizing the following loss (V-REx): 

\begin{equation*}
    \min_{\mathcal{X} \rightarrow \mathcal{Y}} \lambda \cdot \mathrm{Var}\{R^1(\Phi),...,R^{n_e}(\Phi)\} + \frac{1}{|\mathcal{E}_{tr}|} \sum_{e \in \mathcal{E}_{tr}} R^e(\Phi)
\end{equation*}

where $\Phi$ is the invariant predictor, the scalar $\lambda \geq 0$  controls the balance between reducing average risk and enforcing the equality of risks, and $Var$ stands for the variance between the risks across training environments. To the best of our knowledge, these approaches have not been evaluated yet on image segmentation tasks.

\subsection{Domain-Prediction-based methods}
\label{domain_prediction_methods}

Given data from multiple sources, \textbf{Domain-Prediction-based} methods find a harmonized data representation such that all information relating to the source domain of the image is removed \cite{Dinsdale2020.10.09.332973,ganin_unsupervised_2015,martel_dual-task_2020}. 
This goal can be achieved by appending another head to the network, which acts as a domain classifier. During training, the ability of the domain classifier to predict the domain is minimized to random chance, thus reducing the domain-specific information in the data representation. Domain-Prediction-based methods differ from Regularization-based methods in that they rather remove the need to annotate all images from a new target dataset to be able to train a model on new unlabeled data, and so allow leveraging non-annotated data.

As described by Dinsdale et al. \cite{Dinsdale2020.10.09.332973}, the network shown in Fig. \ref{UNet_plot} is a modified U-Net with a second head which acts as a domain predictor. The final goal is to find a representation that maximizes the performance on a segmentation task with input images $\mathbf{X_d} \in \mathbb{R}^{B_d \times W \times H \times D}$ and task labels $\mathbf{Y_d} \in \mathbb{R}^{B_d \times W \times H \times D \times C}$ while minimizing the performance of the domain predictor. This network is composed of an encoder, a segmentor, and a domain predictor with respective weights $\Theta_{repr}$, $\Theta_{seg}$, and $\Theta_{dp}$.

The network is trained iteratively by minimizing the loss on the segmentation task $L_s$, the domain loss $L_d$, and the confusion loss $L_{conf}$. The confusion loss penalizes a divergence of the domain predictor's prediction from a uniform distribution and is used to remove source-related information from $\Theta_{repr}$. It is also important that the segmentation loss $L_s$ be evaluated separately for the data from each dataset to prevent the performance being driven by only one dataset, if their sizes vary significantly. In our work, the segmentation loss $L_s$ takes the form of the sum of a Sorensen-Dice loss and a binary cross-entropy loss. 
The domain loss $L_d$ is used to assess how much information remains in  about the domains. 

The Domain-Prediction method thus minimizes the total loss function:

\begin{align*}
    L(\textbf{X}, \textbf{Y},\textbf{D}, \Theta_{repr}, \Theta_{seg}, \Theta_{dp}) =  &\sum_{d \in \mathcal{E}_{tr}}  L_s(\mathbf{X_d}, \mathbf{Y_d}; \Theta_{repr}, \Theta_{seg}) \\
    &+ \alpha \cdot L_d(\textbf{X}, \textbf{D}, \Theta_{repr};\Theta_{dp}) \\
    &+ \beta \cdot L_{conf}(\textbf{X}, \textbf{D}, \Theta_{dp};\Theta_{repr})
\end{align*}

where $\alpha$ and $\beta$ represent weights of the relative contributions for the different loss functions. $(\mathbf{X_d}, \mathbf{Y_d})$ corresponds to the labeled data available for the domain $d$, while $(\mathbf{X}, \mathbf{D})$ corresponds to all the images (labeled and unlabeled) available and their corresponding domain. This is interesting as training the domain predictor does not require labeled data and semi-supervised training can be used to find a harmonized data representation.

\subsection{A Combined Method for OoD Generalization}

The losses introduced in the context of Regularization-based OoD Generalization can be combined to the Domain-Prediction mechanism. We choose to apply the V-REx loss to the segmentor only. The intuition is that if the segmentor learns robust features through regularization, then there will be less source-related information in the extracted features and can make the Domain-Prediction part of the method easier. Besides, Domain Prediction schemes already require to compute the segmentation loss for each domain separately, which is also a requirement for Regularization-based methods. This method aims to minimize the following loss function:

\begin{align*}
    \min_{\mathcal{X} \rightarrow \mathcal{Y}} 
        \lambda  
        &\cdot \mathrm{Var}\{ 
            L_s(\mathbf{X_1}, \mathbf{Y_1}; \Theta_{repr}, \Theta_{seg}), 
            ..., 
            L_s(\mathbf{X_{n_d}}, \mathbf{Y_{n_d}}; \Theta_{repr}, \Theta_{seg})\} \\
        &+ \sum_{d=1}^{n_d} L_s(\mathbf{X_d}, \mathbf{Y_d}; \Theta_{repr}, \Theta_{seg})\\
            &+ \alpha \cdot L_d(\mathbf{X}, \textbf{D}, \Theta_{repr};\Theta_{dp})\\
            &+ \beta \cdot L_{conf}(\mathbf{X}, \textbf{D}, \Theta_{dp};\Theta_{repr})
\end{align*}

\section{Experimental Setup}

In the following, we present the datasets used in this work and detail the evaluation strategy used.

\subsection{Datasets}

We use a corpus of three datasets for the task of hippocampus segmentation from various studies. The \textit{Decath} dataset \cite{simpson_large_2019} contains MR images of both healthy adults and schizophrenia patients. The \textit{HarP} dataset \cite{boccardi_training_2015} is the product of an effort from the European Alzheimer’s Disease Consortium and Alzheimer’s Disease Neuroimaging Initiative to harmonize the available protocols and contains MR images of senior healthy subjects and Alzheimer's disease patients. Lastly, the \textit{Dryad} dataset \cite{kulaga-yoskovitz_multi-contrast_2015} contains MR images of healthy young adults.

Since both \textit{HarP} and \textit{Dryad} datasets provide whole-head MR images at the same resolution, we crop each scan at two fixed positions, respectively for the right and left hippocampus. The shape of the resulting crops (64, 64, 48) fits every hippocampus in both datasets. Since segmentation instances from the \textit{Decath} dataset are smaller, they are centered and zero-padded to the same shape. Similarly, as all the datasets do not provide the same number of classes in their annotation, we restrict the classes to “background” and “hippocampus”.

\subsection{Evaluation}

We first train three instances of a 3D U-Net, using 3 layers for encoding, and decoding with no dropout, respectively on each of the datasets to assess how well the features learned on a dataset generalize to the other datasets. We then evaluate the proposed method with five-fold cross-validation. In the context of fully supervised training, a model is trained on two of the datasets and evaluated on the third dataset. 10\% of the data in each fold is used for validation and guiding the training schedule. The dataset that is not used to train the model is considered entirely as test data. Due to the nature of the task, we need to find a hyper-parameter/data augmentation configuration that allows the model to generalize well on the test dataset no matter which pair of training dataset is used.

\begin{figure}
\begin{center}
        \includegraphics[width=0.7\textwidth]{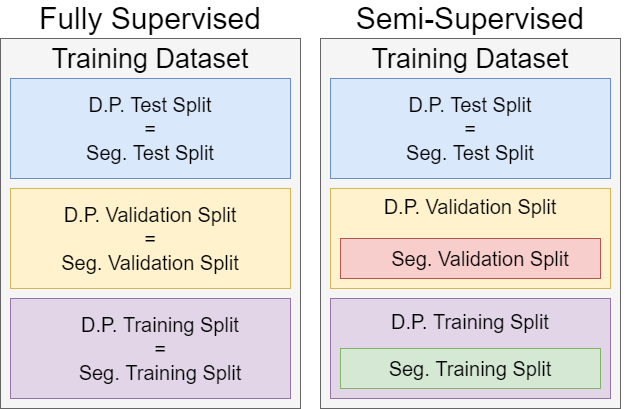}
    \caption{
Schema of testing/validation/training splits in the context of fully supervised and semi-supervised training. The test splits remain equal while the segmentor's training and validation splits are subsets of the splits for the domain predictor. “D.P.” refers to the domain predictor and “Seg.” to the segmentor.
    }
    \label{schema_splits}
\end{center}
\end{figure}

For semi-supervised training, we train our models on all 3 datasets. The splits for the domain predictor are computed the same way as during fully supervised training. As for the segmentor, its training and validation splits are subsets of the domain predictor's respective splits, as shown in Fig. \ref{schema_splits}. 
The test splits for both heads remain equal. The splits are computed only once and reused for each method to reduce the variance during testing.

The network is implemented using Python 3.8 and PyTorch 1.6.0. Some light data augmentation is also used: \textit{RandomAffine}, \textit{RandomFlip}, \textit{RandomMotion}. The batch size for each dataset is selected so that the training duration is the same with all datasets.

\section{Results}

In this section, we first evaluate the generality of the features learned by a standard U-Net on each dataset. 
We then compare the different methods in a fully supervised settings, and we outline their limitations. Finally, we assess whether Domain-Prediction-based methods can leverage their ability of training on unlabeled data.

\subsection{U-Net results on each dataset}
\label{one_dataset_results}

We train three instances of a standard 3D U-Net respectively on each of the datasets, in order to assess how well features learned by a model on a given dataset generalize well to the other two datasets. The results can be seen in \cref{one_dataset}. Each of three trained models achieves a mean Dice score in the range of $84$ to $90$ with a low standard deviation below 2. As a comparison, Isensee et al. \cite{isensee_nnu-net_2018}, Carmo et al. \cite{carmo_hippocampus_2020}, and  Zhu and al. \cite{zhu_dilated_2019} achieve respectively on Decath, HarP, and Dryad a $90$, $86$, and $89$ Dice scores. So, these results are close to state-of-the-art, which is not our goal per se, but they should give a good insight on how features learned in one dataset generalize to the other ones. 

\begin{table}
    \begin{center}
        \begin{tabular}{|c | c c c|}
\hline
 Training dataset & Decath & HarP & Dryad\\
 \hline
Decath \textbf{only} & \bf 88.5 ± 1.4 & 28.7 ± 12.0 & 17.1 ± 14.3\\
HarP \textbf{only} & 76.5 ± 2.4 & \bf 84.6 ± 1.4 & 80.0 ± 1.7\\
Dryad \textbf{only} & 59.0 ± 4.8 & 57.9 ± 1.9 & \bf 85.2 ± 1.9\\
\hline
        \end{tabular}
    \end{center}

\caption{
Dice coefficient, comparing generalization capability of the features learned by the U-Net in each dataset using a standard no OoD Generalization mechanism. The results on the diagonal are for the respective in-distribution dataset of each model and otherwise the results correspond to OoD testing.
}    
\label{one_dataset}
\end{table}

The performance of the three models on OoD Generalization differ greatly. While the model trained on HarP already achieves great results, the other two models seem to struggle much more to generalize and attain a much lower average and higher standard deviation. So, training each of the three pairs of datasets provides a different setting and allows testing multiple scenarios.

\subsection{Fully supervised Training}

For all dataset configuration, we use two datasets for training and the third one for OoD testing. The training datasets are also referred to as in-distribution datasets.

\subsubsection{Training on Decath and HarP:}

In this first setup, we train all models on Decath and HarP. The results can be seen in \cref{decath_harp}. Training a U-Net only on HarP already learns features that generalize well, so we are interested in seeing how adding “lesser quality” data (\cref{one_dataset_results}) influences the results on Dryad. The results are shown in \cref{decath_harp}.
If we first focus on the results on the in-distribution datasets, we observe that all methods achieve state-of-the-art results.
All methods achieve satisfying results on the out-of-distribution dataset with the combined method having a lead. The domain prediction accuracy reaches near random results for all Domain-Prediction-based methods, which indicates that the domain-identifying information is removed.

\begin{table}[h]
    \begin{center}
        \begin{tabular}{|c | c c c| c |}
\hline
Method &  \bf Decath & \bf HarP & Dryad & Dom. Pred. Acc.\\
 \hline
U-Net & 89.5 ± 0.4 & 85.7 ± 0.7 & 81.2 ± 1.5 & -\\
U-Net + V-REx & 89.4 ± 0.6 & 85.3 ± 1.2 & 81.7 ± 0.5 & -\\
Domain-Prediction & \bf 90.0 ± 0.4 & \bf 86.8 ± 1.3 & 82.8 ± 2.5 & 48.6 ± 6.2\\
Combined & \bf 88.9 ± 0.7 & 85.2 ± 0.8 & \bf 84.6 ± 0.4 & 56.1 ± 10.9\\
\hline
        \end{tabular}
    \end{center}

\caption{
Dice coefficients and domain prediction accuracy (when applicable). A \textbf{bold} dataset name denotes that the dataset is used for training.
}     
\label{decath_harp}
\end{table}

\subsubsection{Training on Decath and Dryad:}

\label{fully_sup_decath_dryad}

This training setup is perhaps the most interesting one of the three, as we have seen that training a model on either of Decath and Dryad does not yield good generalization results. As shown in \cref{decath_dryad}, all methods yield state-of-the-art results on both in-distribution datasets. Surprisingly, all methods perform well on HarP, although the regular U-Net to a lesser degree. Regarding domain prediction accuracy, the Domain-Prediction method reaches once more near random chance results. 
The domain prediction accuracy is particularly high for the combined method, since these results stem from an intermediary training stage where the segmentor and domain predictor are trained jointly.

\begin{table}[h]
    \begin{center}
        \begin{tabular}{|c | c c c| c |}
\hline
Method &  \bf Decath & HarP & \bf Dryad & Dom. Pred. Acc.\\
 \hline
U-Net & 88.7 ± 0.7 & 69.8 ± 2.4 & 89.2 ± 0.5 & -\\
U-Net + V-REx & 89.3 ± 0.2 & \bf 73.7 ± 1.9 & 89.9 ± 0.3 & -\\

Domain-Prediction & \bf 89.6 ± 0.4 & 72.9 ± 2.4 & \bf 90.5 ± 0.3 & 56.5 ± 11.5\\

Combined & \bf 89.6 ± 0.2 & 73.2 ± 1.5 & \bf 90.5 ± 0.3 & 90.0 ± 6.3\\
\hline
        \end{tabular}
    \end{center}

\caption{
Dice coefficients and domain prediction accuracy (when applicable). A \textbf{bold} dataset name denotes that the dataset is used for training.
}    
\label{decath_dryad}
\end{table}

\subsubsection{Training on HarP and Dryad:}
\label{fail_dpa}

In this last setup, we train the models on HarP and Dryad and test them on Decath. The results can be seen in \cref{harp_dryad}.
All methods perform almost equally well on in-distribution datasets. However, they differ vastly on the first column, with the U-Net + V-REx method being the clear winner. The regular U-Net and the Domain-Prediction method perform similarly, with the latter having a smaller variance. Removing domain-related information from the U-Net features does not seem to be enough as both Domain-Prediction methods achieve a near random domain prediction accuracy, which indicates that domain-identifying information has been removed, but the combined method vastly underperforms.

\begin{table}[h]
    \begin{center}
        \begin{tabular}{|c | c c c| c |}
\hline
Method &  Decath & \bf HarP & \bf Dryad & Dom. Pred. Acc.\\
 \hline
U-Net & 65.6 ± 18.1 & 84.9 ± 0.8 & 89.8 ± 0.7 & -\\
U-Net + V-REx & \bf 78.1 ± 2.6 & 85.3 ± 0.8 & 90.1 ± 0.3 & -\\
Domain-Prediction & 69.4 ± 6.4 & \bf 86.3 ± 0.8 & \bf 90.6 ± 0.6 & 54.4 ± 7.0\\
Combined & 46.2 ± 10.6 & 84.2 ± 1.1 & 89.7 ± 0.9 & 52.6 ± 9.3\\
\hline
        \end{tabular}
    \end{center}

\caption{
Dice coefficients and domain prediction accuracy (when applicable). The usual data augmentation scheme is used. A \textbf{bold} dataset name denotes that the dataset is used for training.
}    
\label{harp_dryad}
\end{table}

\begin{figure}[h]
    \includegraphics[width=\textwidth]{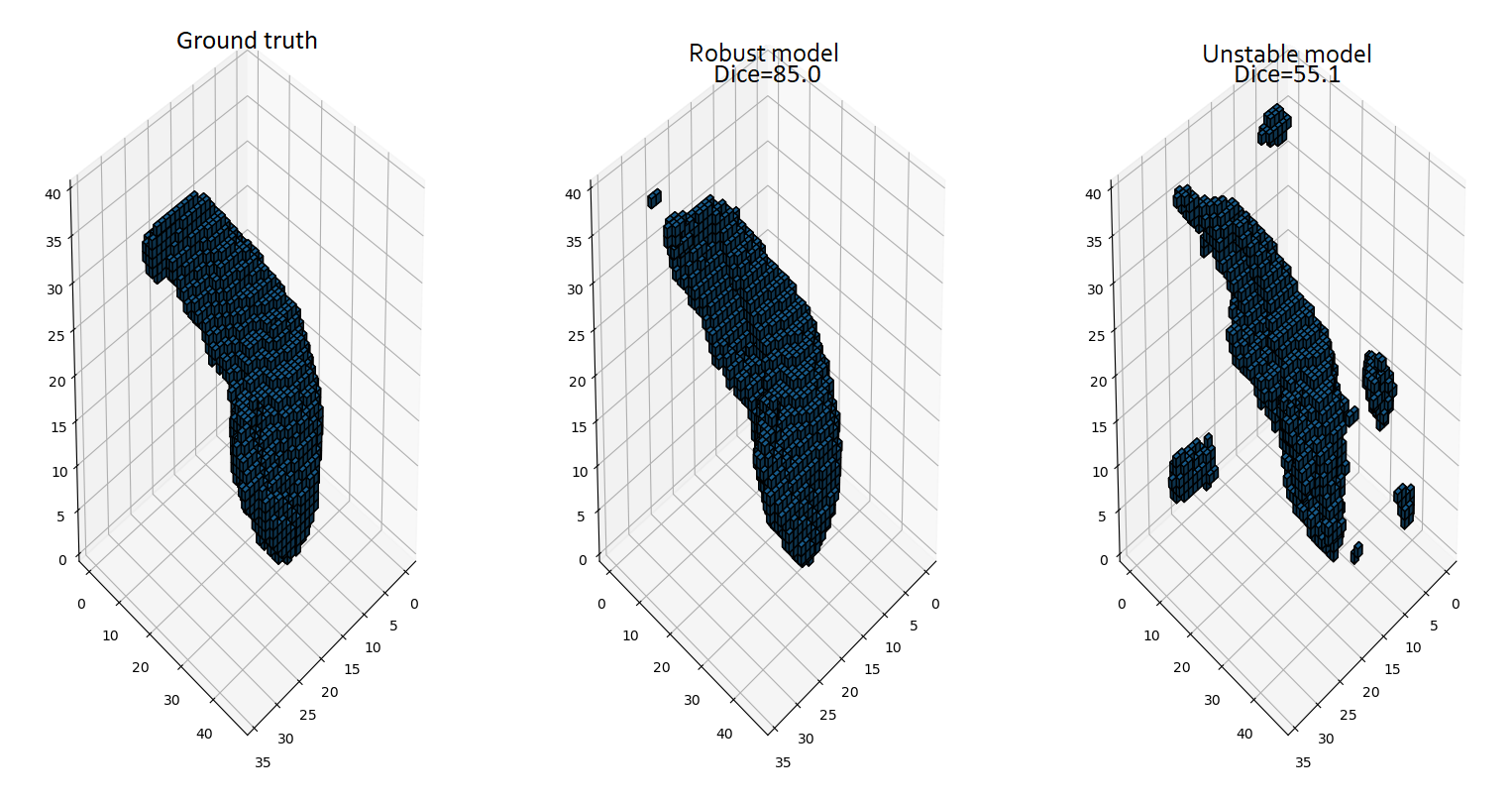}
    \caption{
Segmentation mask prediction for an instance from Decath. Both models are trained using the Domain-Prediction-based method but with different training data folds and achieve respectively $78.9\%$ and $59.9\%$ Dice score on the out-of-distribution dataset Decath. Dice scores on the test splits of the in-distribution datasets are within $1\%$ of each other.
    }
    \label{segmentations_harp_dryad}
\end{figure}

If we observe the segmentation masks predicted by models from different folds of the same cross-validation, we can see in Fig. \ref{segmentations_harp_dryad} that while the first prediction matches the ground truth nicely, the second prediction fails to fully recognize the hippocampus and is very noisy. What is troublesome is that both models achieve Dice scores of $86\%$ and $90\%$ respectively for the test splits of in-distribution datasets. 
The domain prediction accuracy is reduced to random chance without causing any significant drop in performance on the segmentation task. As such, the Domain-Prediction method reduces domain information in the model features as expected.
Reducing domain prediction accuracy does not seem to be enough however, as the combined method also achieves low domain detection results, but fares worse than a regular U-Net on the test dataset. Overall, we have no way of predicting whether a model is usable on new data by looking at the metrics on the validation data. 

It is worth noting that by adding some \textit{RandomBiasField} and \textit{RandomNoise} to the data augmentation scheme, both Domain-Prediction-based methods perform significantly better, as can be seen in \cref{harp_dryad_mri}. The results on the training datasets remain stable with a decrease in standard deviation. The Dice scores on Decath, however, significantly improve. However, this change cannot be justified by just looking at the metrics on validation data. Furthermore, this scheme yields significantly worse results when training on any other dataset pair, which is why it is not used anywhere else.

\begin{table}[h]
    \begin{center}
        \begin{tabular}{|c | c c c| c |}
\hline
Method &  Decath & \bf HarP & \bf Dryad & Dom. Pred. Acc.\\
 \hline
Domain-Prediction & \bf 75.1 ± 1.7 & \bf 84.2 ± 0.9 & \bf 90.0 ± 0.3 & 57.8 ± 7.6\\
Combined & 74.1 ± 4.3 & 82.3 ± 1.2 & 89.0 ± 0.6 & 61.2 ± 3.8\\
\hline
        \end{tabular}
    \end{center}

\caption{
Dice coefficients and domain prediction accuracy (when applicable) for a different data augmentation scheme. A \textbf{bold} dataset name denotes that the dataset is used for training.
}    
\label{harp_dryad_mri}
\end{table}

\subsubsection{Discussion}

OoD Generalization methods seem to be quite limiting in the sense that they require the user to find a hyper-parameter configuration that works well in all setups. 
The results on the validation data are too good to allow us to fine tune the hyper-parameters for each method in each of the settings. 
As such, no method tested gives reliable results, although the U-Net + V-REx method always performs at least on par, and often better than, the reference U-Net.

We also observe that \textbf{we have no way of predicting whether a model is usable on new data by looking at the metrics on the validation data}. The Dice scores on the validation data are not representative of how well a model generalizes on out-of-distribution data. A near random guessing domain prediction accuracy is also not sufficient, as seen in \cref{fail_dpa}. 

\subsection{Semi-supervised Training}

During the fully supervised training, we noted that Domain-Prediction-based methods trained using HarP and Dryad can fail to fully recognize the hippocampus and produce very noisy segmentations. The main goal of these experiments is to test whether adding a few labeled data points from Decath to the training set can alleviate this issue. 

Being able to train on unlabeled data is also a very interesting feature for a method, considering that image annotation is a time-consuming task. 
In the next titles, if present, the numbers between square brackets next to a dataset's name mean refer respectively to the number of labeled data points used for training and validation from this dataset in addition to the unlabeled data points.

\subsubsection{Training on Decath[5,5], HarP and Dryad:} 
\label{Decath55HarpDryad}

In this experiment, we train the models on all datasets but only HarP and Dryad are fully annotated. Decath is almost fully unlabeled. Only 10 labeled points are considered and are split between training and validation.
The results are shown in \cref{decath55_rest}.

\begin{table}[h]
    \begin{center}
        \begin{tabular}{|c | c c c| c |}
\hline
Method &  \bf Decath[5, 5] & \bf HarP & \bf Dryad & Dom. Pred. Acc.\\
 \hline

U-Net & \bf 88.1 ± 0.6 & \bf 85.9 ± 1.0 & \bf 90.5 ± 0.5 & -\\
U-Net + V-REx & 87.8 ± 0.8 & 85.7 ± 1.0 & \bf 90.5 ± 0.4 & -\\
Domain-Prediction & 87.4 ± 0.7 & 85.4 ± 1.4 & 90.0 ± 0.5 & 56.6 ± 5.2\\
Combined & 86.9 ± 1.7 & 85.8 ± 1.1 & 90.3 ± 0.6 & 57.2 ± 1.5\\

\hline
        \end{tabular}
    \end{center}

\caption{
Dice coefficients and domain prediction accuracy (when applicable). A \textbf{bold} dataset name denotes that the dataset is used for training. If present, the numbers between brackets next to an in-distribution dataset's name refer respectively to the number of data points used for training and validation from this dataset.
}    
\label{decath55_rest}
\end{table}

All methods perform similarly on the segmentation task and reach state-of-the-art results on all datasets. The Domain-Prediction-based methods decrease the domain prediction accuracy, but it remains far better than random chance ($55\%$ vs $33\%$). 
During training, sudden variations in metrics for the segmentation tasks were observed  even at learning rates as low as $5 \cdot 10^{-5}$. However, the results achieved here are only slightly worse than the one reached during fully-supervised training. As such, we expect that refining this tuning only leads to marginal gains on the segmentation task. 
Overall, no method here is able to outperform the others and yield significant improvement over using a standard U-Net.

\subsubsection{Training on Decath[6,4], HarP[6,4] and Dryad[6,4]:}

From what we have seen in \cref{Decath55HarpDryad}, training models on two fully labeled datasets and one partially labeled one does not allow discriminating between methods. Instead, we can consider that we have multiple dataset available, but only a few labeled instances for each of them. With only 18 instances being used for training, we want to see whether one method outperforms the others. 

\begin{table}[h]
    \begin{center}
        \begin{tabular}{|c | c c c| c |}
\hline
Method &  \bf Decath[6, 4] & \bf HarP[6, 4] & \bf Dryad[6, 4] & Dom. Pred. Acc.\\
 \hline

U-Net & 86.7 ± 0.7 & 80.2 ± 1.4 & 87.3 ± 1.2 & -\\
U-Net + V-REx & \bf 86.9 ± 0.5 & \bf 80.7 ± 1.0 & \bf 87.5 ± 1.0 & -\\
Domain-Prediction & 82.9 ± 3.3 & 75.7 ± 4.8 & 85.6 ± 1.5 & 56.1 ± 4.4\\
Combined & 85.6 ± 0.5 & 78.4 ± 2.0 & 86.4 ± 0.7 & 54.7 ± 3.7\\

\hline
        \end{tabular}
    \end{center}

\caption{
Dice coefficients label and domain prediction accuracy (when applicable). A \textbf{bold} dataset name denotes that the dataset is used for training. If present, the numbers between parentheses next to an in-distribution dataset's name refer respectively to the number of data points used for training and validation from this dataset.
}    
\label{all64}
\end{table}

As shown in \cref{all64}, all methods perform almost equally well. The results for the U-Net and U-Net + V-REx methods are impressive, considering that there is only a slight drop in performance now that we are training these models using far fewer labeled data points. 
The results for the Domain-Prediction-based methods are slightly worse, and the domain prediction accuracy remains relatively high for both methods. These results are coherent with the ones obtained during the previous experiment as the same hyper-parameter combination is used, although the combined method achieves better results than the standard Domain-Prediction method on the second column.

\subsubsection{Discussion}

Overall, no method reliably outperforms the standard U-Net. Domain-Prediction-based methods actually performs worse than the reference in the last experiment. Either removing domain related information does not scale well with the number of training domains, or a much finer hyper-parameter tuning is required to allow these methods to leverage their ability to train on unlabeled data. The feasibility of this tuning can also be put in question due to the increase in instability observed during training. 

\section{Conclusion}

In this work, we evaluate a variety of methods for \textbf{Out-of-Distribution Generalization} on the task of hippocampus segmentation. In particular, we evaluate \textbf{Regularization-based} methods which regularize the performance of a model across training environments. We also consider a \textbf{Domain-Prediction-based} method, which adds a domain classifier to the architecture, its goal being to reduce the prediction accuracy of this classifier. Lastly, we explore how well a method uniting both approaches performs. 

We compare these methods in a fully supervised setting, and we observe the limitations of OoD Generalization methods. No method performs reliably in all experiments. Only the \textbf{V-REx} loss stands out as it remains easy to tune, while its worst results remains close to the reference U-Net. 

To gauge the ability of Domain-Prediction-based methods to train on unlabeled data, we subsequently evaluate all methods in a semi-supervised settings, using the minimum number of labeled images required to retain a stable training trajectory. The model trained with a V-REx loss maintains impressive results despite the lack of training data, while Domain-Prediction-based methods show their limits when training on an increased number of domains.

The combined method achieves good results in some settings, but suffers from instability in others. Domain-Prediction-based methods have a lot of potential, but require a lot of fine-tuning and can cause the training to slow down.

In the future, we wish to evaluate these methods on another corpus of datasets. This different setting will allow us to evaluate whether the results observed here still hold, especially whether Domain-Prediction-based methods can make use of training on unlabeled data during semi-supervised training. We also wish to confirm whether using the V-REx loss remains a good option in another setting, as it achieves similar or better results than a standard loss on this corpus of datasets.

%
%
%
\bibliographystyle{splncs04}
\bibliography{egbib}

\end{document}